\RequirePackage{lineno}
\documentclass[aps,superscriptaddress]{revtex4-2}
\usepackage{amssymb,amsmath,epsfig}
\usepackage{subfigure}
\usepackage{color}
\usepackage[colorlinks=true, pdfstartview=FitV, linkcolor=blue, citecolor=red, urlcolor=magenta, breaklinks=true]{hyperref}

\usepackage{graphicx}
\usepackage{latexsym}
\usepackage{amsmath}
\usepackage{amsfonts}
\usepackage{amssymb}
\usepackage{booktabs}
\usepackage{multirow}
\usepackage{verbatim}
\usepackage{orcidlink}
\graphicspath{{Graf/}}
\usepackage{array}
\usepackage{float}

\begin{document}
\title{Intermediate Coupling Regime in Dilatonic f(R,T) Inflationary Universe}

\author{F. A. Brito \orcidlink{0000-0001-9465-6868}}\email{fabrito@df.ufcg.edu.br}
\affiliation{Departamento de F\'{\i}sica, Universidade Federal de Campina Grande
Caixa Postal 10071, 58429-900 Campina Grande, Para\'{\i}ba, Brazil}
\affiliation{Departamento de F\'isica, Universidade Federal da Para\'iba, 
Caixa Postal 5008, 58051-970 Jo\~ao Pessoa, Para\'iba, Brazil}

\author{C. H. A. B. Borges}\email{carlos.borges@ifpb.edu.br}
\affiliation{Instituto Federal de Educação Ciência e Tecnologia da Para\'{\i}ba(IFPB), Campus Campina Grande - Rua Tranquilino Coelho Lemos,671,Jardim Dinamérica I, Campina Grande 58432-300, Para\'{\i}ba, Brazil}

\author{J. A. V. Campos \orcidlink{0000-0002-4252-2451}}\email{a.campos@uaf.ufcg.edu.br}
\affiliation{Departamento de F\'{\i}sica, Universidade Federal de Campina Grande
Caixa Postal 10071, 58429-900 Campina Grande, Para\'{\i}ba, Brazil}

\author{F. G. Costa}\email{francisco.geraldo@ifpb.edu.br}
\affiliation{Instituto Federal de Educação Ciência e Tecnologia da Para\'{\i}ba(IFPB), Campus Campina Grande - Rua Tranquilino Coelho Lemos,671,Jardim Dinamérica I, Campina Grande 58432-300, Para\'{\i}ba, Brazil}


\begin{abstract}
In the present work we study cosmology in dilatonic $f(R,T)$ gravity to address the inflationary phase of the early Universe. As usual, in dilatonic gravity the scalar potential assumes the exponential form. However, this potential is not good enough to be in accord with the Planck 2018 data. More strikingly, the generalized $\beta$-exponential cannot take this into account either. It is just the presence of the dilatonic sector, in the intermediate coupling regime, that can help the theory to be in full accord with the observational data. 

\end{abstract}
 \maketitle

\section{Introduction}
The cosmic inflation is a phase suffered by the primordial Universe.  The cosmic inflation theory first proposed by Alan Guth states that in this phase the Universe experienced rapid expansion just after the Big Bang. Such a rapid expansion occurred on a very tiny scale but had great consequences, explaining many cosmological problems, such as the cosmological horizon and the flatness of the universe \cite{guth1981inflationary, albrecht1982cosmology,linde1982new}. Currently, the theory of cosmic inflation is widely accepted as a theory which is strongly supported by cosmological observations, such as the detection of temperature fluctuations in the cosmic microwave background (CMB) radiation, which are consistent with theoretical predictions of the inflationary scenario. Moreover, investigations of distant galaxies and matter distribution in the universe agree with many aspects of the theory \cite{mukhanov1981quantum, guth1982fluctuations, hawking1982development,starobinsky1982dynamics}.

The inflaton field plays an important role in inflationary cosmology. It concerns a hypothetical scalar field that in some aspects is similar to the Higgs field, accepted to cause the phase of rapid expansion \cite{guth1981inflationary}. This field is governed by a scalar potential that, at the limit it is predominant, can develop the exponential expansion of spacetime. In particular, the potential $V(\phi)$ of the inflaton field $\phi$ describes the potential energy associated with different configurations of the inflaton. Thus, a specific form of this potential is of great importance in determining the characteristics of inflation, as for example its duration and scale of expansion. Many forms of the inflaton potential have been put forward, since those including quadratic potentials, plateau-like potentials, to more complex forms \cite{hossain2014class, martin2014best, martin2014encyclopaedia,geng2015quintessential, huang2016inflation}

The main common feature of the inflaton potential is to support a phase of rolling inflaton field from high to low potential energy during inflation. Thus, as the field rolls from the top to the bottom of the potential, such that the potential energy is dominant in this phase, yields an accelerated expansion of the universe. The inflationary phase ends as the inflaton field approaches the minimum of its potential and then starts oscillating. Such oscillations convert the inflaton energy to other particles and fields. This process is called ``reheating", which heats the universe and initiates the radiation phase \cite{linde2005particle, dolgov1982baryon, abbott1982particle,starobinsky1984nonsingular,dolgov1999equation,traschen1990particle}.

The cosmic inflation theory has been extended to various modified gravity scenarios, which have $f(R)$, $f(G)$, and $f(\mathcal{T})$ theories as good examples. As now usual, in $f(R)$ gravity, the Ricci scalar $R$ in the Einstein-Hilbert action is formally replaced by a general function $f(R)$. This modification naturally lead to an inflationary phase of the universe since it introduces additional degrees of freedom via the function $f(R)$, which can lead to an accelerated expansion of the early universe without introducing an inflaton field  \cite{odintsov2019f, oikonomou2018exponential, kleidis2018scalar, nojiri2017constant}. Similarly, gravity $f(G)$, that is, the Gauss-Bonnet invariant $G$ is generalized to a function $f(G)$. As is well known, the Gauss-Bonnet term is a topological term in four-dimensional spacetime, and as a consequence it does not affect the equations of motion in standard general relativity. However, in extended form, $f(G)$ can contribute to the dynamics of the universe, providing a mechanism for inflation. Thus, including $f(G)$ terms can naturally produce a richer variety of inflationary solutions, potentially addressing issues such as the so-called graceful exit problem and the generation of primordial perturbations \cite{zhong2018inflation}. In the same perspective, $f(\mathcal{T})$ gravity makes an interesting modification as to lead to teleparallel gravity by replacing the torsion scalar $\mathcal{T}$ with a function $f(\mathcal{T})$ \cite{rezazadeh2017logamediate}. This theory does not include the curvature of spacetime, but instead uses torsion to describe gravitational interactions. Finally, in a more strict way $f(\mathcal{T})$ gravity can lead to an inflationary epoch by appropriately choosing the form of $f(\mathcal{T})$, leading to accelerated expansion. These models have been considered in the recent literature to show that they produce viable inflationary scenarios consistent with observational data, i.e., they survive under constraints from the cosmic microwave background and large-scale structure.

The $f(R,T)$ gravity is a modification of general relativity that extends the $f(R)$ gravity by introducing a dependence on the trace of the energy-momentum tensor $T$. This addition leads to a theory that accounts for matter and geometry coupling, by introducing new dynamical aspects that affect the cosmic evolution \cite{shabani2014cosmological, moraes2017modeling, bhattacharjee2020inflation,simony1}. The $f(R,T)$ theory has been recently applied to cosmic inflation, shown promising results, such as the viability of the theory against various cosmological observations. In such a scenario, the $f (R,T)$ gravity is coupled to an inflaton field.

A crucial field of superstring models and of cosmological scenarios based on the string effective action, is the dilaton field. This field drives the effective strength of all gauge couplings in ``grand-unified" models of fundamental interactions \cite{witten1984some}. Such coupling strengths can lead the Universe towards a strong coupling phase, preceding the phase of standard decelerated evolution. In $M$-theory, the dilaton is naturally interpreted as the effective ``radius" of the eleventh dimension \cite{witten1995string}. This field may affect inflationary dynamics and play a fundamental role in generating the primordial spectra of quantum fluctuations, which can be amplified by inflation. As such, the dilaton could naturally provide the coupled quintessence scenario, as long as the cosmological evolution of this field continues after entering the weak coupling regime. Such a regime characterized by very light (or massless) dilaton is weakly coupled to matter \cite{gasperini2001dilatonic}. On the other hand, massive dilaton is more strongly coupled to macroscopic matter from the gravitational point of view. A sector that was studied in detail in \cite{gasperini2001quintessence}.

One should notice that the coupling constant is related to the dilaton field through the expression $g_{s}=e^{\phi}$. Thus, the weak and strong coupling limits are given by $\phi\longrightarrow-\infty$ and $\phi\longrightarrow\infty$, respectively. Recently, the authors studied $f(R, T)$ modified theories of gravity in the context of string theory inspired dilaton gravity \cite{brito2024weak} in the weak coupling limit. It was considered a specific model that under certain conditions describes the late time Universe in accord with recent observational data. {In doing so, we employed numerical methods to obtain several important observable quantities.}

On the other hand, the intermediate regime in which $\phi$ moves between these two extremes, remains little explored. Thus, in the present study, we shall focus on this limit to address inflationary cosmology in dilatonic $f(R,T)$ gravity.

The paper is organized as follows.  {In Sec.~\ref{III}, we review the formalism of pure $f(R,T)$ gravity and introduce the dilatonic $f(R,T)$ gravity by explicitly informing the role of the coupling to be studied in the intermediate regime.} In Sec.~\ref{IV} we explicitly address the equation of motion for the dilaton playing the role of an inflaton field. In Sec.~\ref{V} we focus on the inflationary parameters and show our main results. Finally in Sec.~\ref{VI} we make our final considerations.


\section{Cosmology in dilatonic $f(R,T)$ Gravity}\label{III}

Let us first take the action given in \cite{harko2011f}
\begin{equation}
    S=\frac{1}{2\kappa}\int d^{4}x f(R,T)\sqrt{-g}+\int d^{4}x{\cal L}_{m}\sqrt{-g},
    \label{1.1}
\end{equation}
where $f(R,T)$ is an arbitrary function of the Ricci scalar curvature $R=g^{\mu\nu}R_{\mu\nu}$ and $T=g^{\mu\nu}T_{\mu\nu}$ is the trace of the energy-momentum tensor, ${\cal L}_{m}$ is the Lagrangian density of matter and $\kappa=8 \pi G$.

Admitting the definition of the energy-momentum tensor, we can express it in such a way that the Lagrangian density of matter depends only on $g_{\mu\nu}$, that is,
\begin{equation}
    T_{\mu\nu}=g_{\mu\nu}{\cal L}_{m}-2\frac{\partial{\cal L}_{m}}{\partial g^{\mu\nu}}.
    \label{key}
\end{equation}
By varying the action given in equation \eqref{1.1} in relation to $g^{\mu\nu}$, we have the field equations given by
\begin{eqnarray}
    f_{R}(R,T)R_{\mu\nu}&+&g_{\mu\nu}\Box f_{R}(R,T)-\nabla_{\mu}\nabla_{\nu}f_{R}(R,T)\nonumber\\
    &+&f_{T}(R,T)(T_{\mu\nu}+\Theta_{\mu\nu})-\frac{1}{2}f(R,T)g_{\mu\nu}-8\pi T_{\mu\nu}=0,	
\end{eqnarray}
so that $(T_{\mu\nu}+\Theta_{\mu\nu})$ corresponds to the variation of the trace with respect to the metric tensor, with $ \Theta_{\mu\nu}\equiv g^{\alpha\beta}\;\frac{\delta T_{\alpha\beta}}{\delta g^{\mu\nu}} $ set in \cite{harko2011f}. We will denote $f_{R}(R,T)$ and $f_{T}(R,T)$ being the derivatives of $f(R,T)$ with respect to the Ricci scalar curvature and the trace of the energy-momentum tensor, respectively.

From the definition of $\Theta_{\mu\nu}$ and using equation \eqref{key}, we have
\begin{equation}
    \Theta_{\mu\nu}=-2T_{\mu\nu}+g_{\mu\nu}{\cal L}_{m}-2g^{\alpha\beta}\frac{\partial^{2}{\cal L}_{m}}{\partial g^{\mu\nu}\partial g^{\alpha\beta}}.
    \label{theta tensor}
\end{equation}
In other words, $\Theta_{\mu\nu}$ will depend on the Lagrangian of matter, that is, this can refer to the case of the electromagnetic field, the massless scalar field, the case of the perfect fluid, among others.


Let us now consider cosmological scenarios related to the effective action that comes from low energy string theory in which the dilaton field exerts influence in the dynamics of the Universe. We shall focus on the sector of the effective action, coming from a low energy string theory, given by the tensor field (the metric $\tilde{g}_{\mu\nu}$) and a scalar field (the dilaton $\phi$), where the tilde indicates that we are working on the \textit{string frame}, the dilaton couples to the Ricci scalar and dilatonic dynamics in an explicit form --- See below.
Our starting point is the string-frame, low-energy, gravidilaton effective action, to lowest order in the $\alpha'$ expansion, but including dilaton-dependent loop and non-perturbative corrections, encoded in a few ``form factors’’, due to the loop corrections $\psi(\phi)$ and $Z(\phi)$ \cite{damour1994string}. $V(\phi)$ is the effective dilaton potential. The model action is \cite{gasperini2001dilatonic}
\begin{equation}
    S=-\frac{M_{P}^{2}}{2}\int d^{4}x\sqrt{-\tilde{g}}\left[e^{-\psi(\phi)}\tilde{R}+\tilde{Z}(\phi)\left(\tilde{\nabla}\phi\right)^2+\frac{2}{M_{P}^{2}}\tilde{V}(\phi)\right]+ \tilde{S}_{\rm{m}}(\phi,\tilde{g},\rm{matter}).
    \label{sframe-1}
\end{equation}
In the cases analyzed in the literature, one can discuss the phenomenology of the relic dilaton background by taking into account two possibilities. First, massive dilaton is gravitationally more \textit{strongly coupled} to macroscopic matter. In strong coupling limit $\phi\to\infty$ we assume that is possible to make an asymptotic Taylor expansion in inverse powers of the coupling constant $g_s^2=\exp(\phi)$ similar to what is done in the context of the ``induced gravity”. Second, massless dilaton is weakly coupled to matter, that we shall
disregard.

We can now characterize the dynamical evolution of the Universe with a metric minimally coupled to the dilaton. In this frame the string effective action is also minimally coupled to perfect fluid sources. Considering the lowest order $\alpha'$ we have \cite{gasperini2001dilatonic}. We can use a more convenient coordinate system so called \textit{Einstein frame} in terms of the metric $g_{\mu\nu}$ that is defined by a conformal transformation $\tilde{g}_{\mu\nu}=e^\phi g_{\mu\nu}$. In this frame, and assume that the action in equation \eqref{sframe-1} depends not only on $R$, but on a function $f(R,T)$, we can rewrite this gravidilaton action as
\begin{equation}
    S=-\frac{M_{P}^{2}}{2}\int d^{4}x\sqrt{-g}\left[f(R,T)-\frac{1}{2}k^2(\phi)\left(\nabla\phi\right)^2+\frac{2}{M_{P}^{2}}V(\phi)\right]+ S_{\rm{m}}(\phi,e^\phi g_{\mu\nu},\rm{matter}). 
    \label{eframe}
\end{equation}
In this equation $R$ is the Ricci scalar curvature and $T$ is the trace of the energy-momentum tensor of the dilatonic field. In what follows we will also consider $M^2_P = 1/8\pi G = 2$, except otherwise indicated. We have defined
\begin{equation}
    V = e^{2\psi(\phi)}\tilde{V},
    \label{new pot}
\end{equation}
\begin{equation}
    k^2(\phi)=3\psi'^2(\phi)+2Z(\phi)e^{\psi(\phi)}.
    \label{def.k}
\end{equation}
By varying equation \eqref{eframe} with respect to $g_{\mu\nu}$, we will obtain
\begin{equation}
    f_{R}(R,T)R_{\mu\nu}-\frac{1}{2}f(R,T)g_{\mu\nu}+(g_{\mu\nu}\Box-\nabla_{\mu}\nabla_{\nu})f_{R}(R,T)=\frac{1}{2} T_{\mu\nu}-f_{T}(R,T)T_{\mu\nu}-f_{T}(R,T)\Theta_{\mu\nu},
    \label{modifield eistein eq1}
\end{equation}
where {$T_{\mu\nu}=T_{\mu\nu}^{f}+T_{\mu\nu}^{\phi}$} is total energy-momentum tensor of the perfect fluid ($f$) that fills the Universe (baryonic matter, radiation and dark matter) plus the dilaton field. Using equation \eqref{theta tensor}, we can write
\begin{eqnarray}
    \Theta_{\mu\nu}=-2\left(T_{\mu\nu}^{\phi}+T_{\mu\nu}^{f}\right) +g_{\mu\nu}\left( {\cal L}^{\phi}+{\cal L}^{f}\right),
    \label{eq theta 1}
\end{eqnarray}
with ${\cal L}^{\phi}$ and $T_{\mu\nu}^{\phi}$  given by
\begin{eqnarray}
    {\cal L}^{\phi}=\frac{1}{2}k^2(\phi)(\nabla\phi)^2-V(\phi),
    \label{lagrangian}
\end{eqnarray}
\begin{eqnarray}
    T^{\phi}_{\mu\nu}=-\frac{2}{\sqrt{-g}}\frac{\delta(\sqrt{-g}{\cal L}^{\phi})}{\delta g^{\mu\nu}}=g_{\mu\nu}{\cal L}^{\phi}-2\frac{\partial {\cal L}^{\phi}}{\partial g^{\mu\nu}}.
    \label{tem}
\end{eqnarray}
In general $f(R,T)$ models, the stress-energy tensor of matter is not covariantly conserved. As a consequence test particles moving in a gravitational field do not follow geodesic lines. There are, however, certain specific forms of the function $f(R,T)$ in which the standard conservation is present. However, we note that such a function $f(R,T)$ must take the particular form $f(R,T)=f_1 (R)+f_2 (T)$. In other words, there is no mixing involving both dependences on $R$ and $T$ \cite{T1T}. Thus, we can think of the purely $T$-dependent term as a redefinition of the matter sector in a minimally coupled gravity theory \cite{T2T}, which illustrates the fact that the coupling of matter and energy with gravity interferes with the conservation law to be obeyed by the energy-momentum tensor. Constraints on the function $f_2 (T)$ were studied in \cite{T3T}. Recently, some works have explored conservation problems in various contexts of application of $f(R,T)$ theories  \cite{T4T}. In our present study we shall consider the previously mentioned form which yields to energy-momentum conservation.

We assume that the energy-momentum tensor of the matter and energy is given by $T_{\mu\nu}=(\rho+p)u_{\mu}u_{\mu\nu}-pg_{\mu\nu}$, with conditions $u_{\mu}u^{\mu}=1$ and $u^{\mu}\nabla_{\nu}u_{\mu}=0$ satisfied by the four-velocity $u^{\mu}$. 
In the following we will consider a simple example of $f(R,T)$ theories in order to study the inflationary phase of the Universe with the presence of a single dilatonic field.

\section{Inflation in dilatonic $f(R,T)$ gravity}\label{IV}

We are interested in studying the dynamics of the inflaton field in the absence of ordinary matter. In other words, we are only interested in the primordial evolution of the Universe dominated by the energy contained in a single scalar field represented by the dilaton field minimally coupled to the gravitational field. So let us introduce a model for an inflationary scenario and obtain the expressions of the inflation observables. Let us now assume a homogeneous and isotropic Universe described by the Friedmann-Lemaitre-Robertson-Walker
(FLRW) metric whose line element is written as 
\begin{equation}
    ds^2=dt^2-a^2(t)(dr^2+r^2d\theta^2+r^2\sin^2(\theta)d\phi^2).
    \label{metric}
\end{equation}
The specific $f(R,T)$ model is $f(R,T)=R+\alpha T$ which is the same covered in \cite{brito2024weak} and was first studied by Harko et al. in \cite{harko2011f}. For the dilatonic $f(R,T)$ gravity model \eqref{eframe}, the field equations \eqref{modifield eistein eq1}, with the Lagrangian \eqref{lagrangian}, can be written as
\begin{equation}
    G_{\mu\nu}=\left( \frac{1}{2}-\alpha\right)T_{\mu\nu}-\alpha\Theta_{\mu\nu}+\frac{1}{2}g_{\mu\nu}\alpha T,
    \label{key-something}
\end{equation}
with
\begin{eqnarray}
T_{00}&=&\frac{1}{2}k^{2}(\phi)\dot{\phi}^{2}+V(\phi),\\
T_{ii}&=&\frac{1}{2}k^{2}(\phi)a^{2}\dot{\phi}^{2}-a^{2}V(\phi),\\
T&=&-k^{2}(\phi)\dot{\phi}^{2}+4V(\phi).
\end{eqnarray}
The components of the tensor $\Theta_{\mu\nu}=-2T_{\mu\nu}+g_{\mu\nu}{\cal L}$ are given by
\begin{eqnarray}
\Theta_{00}&=&-\frac{1}{2}k^{2}(\phi)\dot{\phi}^{2}-3V(\phi)\\
\Theta_{ii}&=&-\frac{3}{2}k^{2}(\phi)a^{2}\dot{\phi}^{2}+3a^{2}V(\phi).
\end{eqnarray}
Thus, the field equations are
\begin{equation}
    3H^{2}=\left( \frac{1}{2}-\alpha \right) \frac{1}{2}k^{2}(\phi)\dot{\phi}^{2}+\left( \frac{1}{2}+4\alpha \right)V(\phi),
    \label{friedmann1}
\end{equation}
\begin{equation}
    2\dot{H}+3H^{2}=-\left( \frac{1}{2}+3\alpha\right)\frac{1}{2}k^{2}(\phi)\dot{\phi}^{2}+\left( \frac{1}{2}+4\alpha \right)V(\phi).
    \label{friedmann2}
\end{equation}
The dilaton equation of motion is given by
\begin{equation}
    (1-2\alpha)k^{2}(\phi)\ddot{\phi}+3(1+2\alpha)k^2(\phi)H\dot{\phi}+(1-2\alpha)k(\phi)k^{\prime}(\phi)\dot{\phi}^{2}+(1+8\alpha)V^{\prime}(\phi)=0, 
    \label{dil.equation}
\end{equation}
where ``dot'' denotes derivative with respect to time. From equations \eqref{friedmann1} and \eqref{friedmann2} we have
\begin{equation}
    \dot{H}=-\frac{1}{4}(1+2\alpha)k^2(\phi)\dot{\phi}^2.
    \label{Hponto}
\end{equation}
Let us use the well-known slow-roll approximation, which states a
dynamic regime $ \dot{\phi}^{2}\ll V(\phi) $ and $ \ddot{\phi}\ll\dot{\phi}$. Thus, in this regime, the equations \eqref{friedmann1} and \eqref{dil.equation} can be written as
\begin{equation}
    H^{2}=\frac{1}{6}(1+8\alpha)V(\phi),
    \label{key2}
\end{equation}
\begin{equation}
    k^{2}(\phi)H\dot{\phi}=-\frac{(1+8\alpha)V^{\prime}(\phi)}{3(1+2\alpha)},
    \label{key3}
\end{equation}
{as long as $\frac12 k^2(\phi)\dot{\phi}^2\ll V(\phi)$ and consequently $k^2(\phi)\ddot{\phi} +k(\phi)k'(\phi)\dot{\phi}^2\ll V'(\phi)$}.
In this way, the dilatonic field, which plays the role of the inflaton field in this theory, can be written as
\begin{eqnarray}
    \dot{\phi}=-\frac{\sqrt{6(1+8\alpha)}V^{\prime}(\phi)}{3k^{2}(\phi)(1+2\alpha)\sqrt{V(\phi)}}.
    \label{final.dil.eq}
\end{eqnarray}
{This equation can still be written as 
\begin{equation}
    \frac{k^2(\phi)\dot{\phi}^2}{V(\phi)}=\frac{2(1+8\alpha)V'(\phi)^2}{3k^2(\phi)(1+2\alpha)^2 V^2(\phi)}\propto \epsilon_1.\nonumber
\end{equation}
Notice that $k^2(\phi)\dot{\phi}^2\ll 2V(\phi)$ is ensured because is proportional to the spectral index $\epsilon_1$ (see below). And this is in accord with being very small in the slow-roll regime, where $\epsilon_1\ll1$.}

In the context of the Gasperini-Veneziano theory, the transition between the pre-Big Bang scenario and the post-Big Bang era is marked by a complex dynamics. Before the Big Bang, the universe exists in a state of high density and energy, where the dilaton field plays a crucial role \cite{gasperini2003pre, gasperini2008dilaton}. In this regime, the coupling constant $g_s=e^{\phi/2}$ is very large, reflecting strong interactions among particles. Thus, we can approximate the form factor to  $k^2(\phi)=2c_2^2/c_1^2$ \cite{gasperini2008dilaton}. As the universe expands, an abrupt transition occurs and the inflation field is driven by the dilaton field, which, at the beginning of inflation, takes on very high values \cite{linde1983chaotic}. This elevation in the inflaton (dilaton) field, is necessary to provide the potential energy that triggers the rapid expansion of the universe. During this process, $g_s$ decreases significantly as the energy density decreases, allowing the universe to stabilize \cite{gasperini2008dilaton}. After inflation, the value of the inflaton field reduces, resulting in a lower $g_s$, which facilitates the formation of structures and the dynamics of radiation and matter. This transition is essential for understanding the evolution of the cosmos, connecting the extreme states of the pre-Big Bang to the formation of the observable universe we know today, where the nature of fundamental interactions is profoundly altered. Since we are specifically investigating the impact of a modified theory of gravity on inflationary dynamics and how the dilaton field affects this inflationary scenario, as well as how the inflationary parameters will behave in light of recent experimental data, we will adopt an intermediate form for the term $k^2(\phi)$, which preserves the asymptotic values in the weak and strong field limits. Thus, we use the asymptotic form for the general expression of the form factor \cite{gasperini2008dilaton}
\begin{eqnarray}
k^{2}(\phi)= 2\dfrac{c_{2}^{2}}{c_{1}^{2}}\left[1 - \left(\dfrac{b_{2}}{c_{2}^{2}} + \dfrac{b_{1}}{c_{1}^{2}}\right)\exp[-\phi]\right]. 
\end{eqnarray}
{The constants $c_1, c_2$ and $b_1, b_2$ are dimensionless arbitrary constants that come from combinations of original constants from other form factors.}
The choice for the potential scalar field $V(\phi)$ is very important and characterizes the model we are dealing with. A potential predicted by particle physics is known as the Coleman-Weinberg potential, which arises in the context of grand unified theories \cite{linde1982new, coleman1973radiative}. Other possibilities are, for example, simpler options such as $V(\phi)\sim\phi^2$ or $V(\phi)\sim\phi^4$. This approach is likely, since we do not know which theory of particle physics that describes the early universe. These potentials are considered in detail in reference \cite{belinsky1985inflationary}. Potentials of the type $V(\phi)\sim\phi^4$ were initially considered in the context of chaotic inflation \cite{linde1983chaotic}. Recently, the potential of the type $V(\phi)\sim\phi^2$ has been used to study inflation in the context of the $f(R,T)$ theories \cite{bhattacharjee2020inflation}. Another simple and interesting model is based on the usual exponential function given by exp$(-\lambda\phi)$, as originally investigated by Peebles and Ratra \cite{peebles1988cosmology, ratra1988cosmological, wetterich1994cosmon, ferreira1998cosmology}. Such exponential potentials arise in four-dimensional effective theories induced by Kaluza-Klein theories \cite{lucchin1985power}. Other examples include supergravity and superstring theories \cite{halliwell1987scalar} or theories with high derivative terms \cite{wetterich1985kaluza, shafi1985inflation}. Although conventional inflation cannot be described by the exponential potential, there is a solution with power low inflation and, moreover, this solution is an attractor, in the sense of dynamical systems theories \cite{halliwell1987scalar}.

A possible generalization to potential inflation was introduced in \cite{alcaniz2007beta} and is given by
\begin{eqnarray}
\label{potexp}
V(\phi) & = & V_0 \exp_{1-\beta}{\left(-\lambda\phi/M_{pl}\right)} \nonumber \\ & = &V_0\left[1 + \beta \left(-\lambda \phi/M_{pl}\right) \right]^{1/\beta}\;,
\end{eqnarray}
named $\beta$-exponential potential. 

The $\beta$-exponential function, $\exp_{1-\beta}{({f})}$, is defined as above for positive values of the term between brackets and zero otherwise,  and satisfies the inverse identity $\exp_{1-\beta}\left[\ln_{1-\beta}({f}) \right] =  {f}$, where $\ln_{1-\beta}({f}) = (f^{\beta} - 1)/\beta$ is the $\beta$-logarithmic function \cite{abramowitz1968handbook, lima2001nonextensive}. In the limit $\beta \rightarrow 0$ all the above expressions reproduce the usual exponential and logarithm properties. Actually, $\beta$-exponential potentials present a number of cosmological solutions for a large interval of values of $\beta$. As we will see in the next section, differently from the conventional result, one can also compute the value of the field at the end of inflation using a \textit{slow-roll} parameter.



%


\section{Inflationary Parameters}\label{V}

The gravity theories aimed at explaining cosmic inflation seek to provide estimates that are consistent with observations for different parameters related to inflation. These parameters include: the scalar spectral index of curvature perturbations $n_s$, the ratio of tensor to scalar perturbations $r$ and the tensor spectral index $n_T$ \cite{benisty2020scale}. Observational data impose stringent constraints on these quantities. Generally, these observables are analyzed under the slow-roll hypothesis, where all relevant information about any inflationary scenario is encoded in the slow-roll parameters.  
{In the following steps, we will recover the reduced Planck's mass $M_{pl}=1/\sqrt{8\pi G} \cong 2.44\times 10^{18}GeV$, thus, the equations \eqref{key2} and \eqref{final.dil.eq} can be written respectively as:
\begin{eqnarray}
H^{2} =\dfrac{(1+4\alpha M_{pl}^{2})V(\phi)}{3M_{pl}} \qquad \text{and} \qquad \dot{\phi} = -\dfrac{ V'(\phi)M_{pl}\sqrt{1+4\alpha M_{pl}^{2}}}{(1+\alpha M_{pl}^{2})k^{2}(\phi)\sqrt{3 V(\phi)}}.
\end{eqnarray}
}
Particularly, one may introduce $\epsilon_{n+1}=\frac{d}{dN}$ln$\vert\epsilon_n\vert$, which leads us to $r=16\epsilon_1$, $n_s=1-2\epsilon_1-\epsilon_2$ and $n_T=-2\epsilon_1$ \cite{martin2014encyclopaedia}.
The first two terms of $\epsilon_n$ read \cite{liddle2000cosmological, lyth1999particle}, written in terms of the potential \eqref{potexp} and its derivatives as 


{
\begin{eqnarray}
\epsilon_{1} &\equiv& -\frac{\dot{H}}{H^{2}} = \frac{M_{pl}^{2}}{2k(\phi)^2(1+\alpha M_{pl}^{2})}\left(\frac{V'(\phi)}{V(\phi)}\right)^{2},\\
\epsilon_{2} &\equiv& -\frac{2\dot{H}}{H^{2}}+\frac{\ddot{H}}{H\dot{H}} =\frac{2M_{pl}^{2}}{k(\phi)^2(1+\alpha M_{pl}^{2})} \left[\left(\frac{V'(\phi)}{V(\phi)}\right)^2-\frac{V''(\phi)}{V(\phi)} +\dfrac{V'(\phi)k'(\phi)}{V(\phi) k(\phi)} \right].
\label{param_c}
\end{eqnarray}}

For the potential adopted in the model, we can express the parameters in terms of $\phi$ as
{
\begin{eqnarray}
\epsilon_1 &=& \dfrac{c_{1}^{2}\lambda^{2}}{4c_{2}^{2}(1+\alpha M_{pl}^{2})\left[1-\left(\dfrac{b_{2}}{c_{2}^{2}} + \dfrac{b_{1}}{c_{1}^{2}}\right)e^{-\phi/M_{pl}}\right]\left(1-\dfrac{\beta \lambda \phi}{M_{pl}}\right)^{2}},  \\
\epsilon_2 &=& \dfrac{c_{1}^{2}\lambda^{2}}{c_{2}^{2}(1+\alpha M_{pl}^{2})\left(1-\dfrac{\beta \lambda \phi}{M_{pl}}\right)^{2}\left[1-\left(\dfrac{b_{2}}{c_{2}^{2}} + \dfrac{b_{1}}{c_{1}^{2}}\right)e^{-\phi/M_{pl}}\right]^{2}}\left[\beta - \dfrac{(1+\beta \lambda(2-\phi/M_{pl}))}{2\lambda}\left(\dfrac{b_{2}}{c_{2}^{2}} + \dfrac{b_{1}}{c_{1}^{2}}\right)e^{-\phi/M_{pl}}\right]
\label{param_d}
\end{eqnarray}}
The end of the inflationary phase is expected to occur at $\phi_{e}$  where the condition $\epsilon_{1}(\phi_{e}) \approx 1$ is satisfied, i.e.,
\begin{eqnarray}
\dfrac{\lambda^{2}c_{1}^{2}}{4c_{2}^{2}(1+\alpha M_{pl}^{2})} \approx \left[1-\left(\dfrac{b_{2}}{c_{2}^{2}} + \dfrac{b_{1}}{c_{1}^{2}}\right)e^{-\phi_{e}/M_{pl}}\right]\left(1-\dfrac{\beta \lambda \phi_{e}}{M_{pl}}\right)^{2}.
\label{phi_e}
\end{eqnarray}
We can obtain the scalar field at the end of inflation $\phi_{e}$, by solving numerically the above equation.
We can also write the equations that describe the model in terms of the number of e-foldings $N$, defined by $N\equiv$ 
ln$(a/a_i)$, where $a_i$ represents the scale factor at the beginning of inflation. Using expansion to $k(\phi)$, the definition for $N$ and the potential \eqref{potexp} we can write
\begin{eqnarray}
N &=& \int^{\phi_{e}}_{\phi_{i}}\frac{H}{\dot{\phi}}d\phi=-\frac{(1+\alpha M_{pl}^{2})}{M_{pl}^{2}}\int^{\phi_{e}}_{\phi_{i}}k^2(\phi)\frac{V(\phi)}{V'(\phi)}d\phi,\nonumber \\
&=&\dfrac{(1+\alpha M_{pl}^{2})c_2^2}{c_1^2\lambda}\left[2\left(\dfrac{b_{2}}{c_{2}^{2}} + \dfrac{b_{1}}{c_{1}^{2}}\right)\left(1-\beta \lambda\left(1+\dfrac{\phi}{M_{pl}}\right)\right)e^{-\phi/M_{pl}}- \dfrac{\phi}{M_{pl}}\left(\dfrac{\beta\lambda \phi}{M_{pl}} - 2\right)\right]_{\phi_{i}}^{\phi_{e}},
\label{dil_N}  
\end{eqnarray}
where $\phi_i$ is the value of the dilaton that emerges from the end of the pre-Big Bang phase and assumes the initial value of the scalar field in the inflationary period. 

We can obtain the amplitude of the potential \eqref{potexp} from the primordial power spectrum of scalar perturbations
\begin{eqnarray}
P_{R} = \dfrac{H^{2}}{8\pi^{2}\epsilon_{1}M_{pl}^{4}} \qquad \Rightarrow \qquad P_{R} = \dfrac{(1+4\alpha M_{pl}^{2})V(\phi)}{24\pi^{2}\epsilon_{1}M_{pl}^{4}}\bigg|_{k=k_{*}}
\label{potspetro}
\end{eqnarray} 
where $k_{*}$ is the pivot scale, at which the CMB crosses the Hubble horizon during inflation, its value is fixed at $k_{*} = 0.05Mpc^{-1}$. We also have that the value of $P_{r}(k_{*})$ is fixed by Planck Collaboration (2018) $P_{r}(k_{*}) = A_{s} = 2.0933\times10^{-9}$, so from \eqref{potspetro} we can express the amplitude as follows
\begin{eqnarray}
V_{0} = \dfrac{24\pi^{2}\epsilon_{1}(\phi_{i})M_{pl}^{4}}{(1+4\alpha M_{pl}^{2})\left(1-\beta \lambda \phi_{i}/M_{pl}\right)^{1/\beta}}.
\label{Amplpot}
\end{eqnarray} 

\begin{figure}[H]
\centering
\subfigure[]{\includegraphics[scale=0.6]{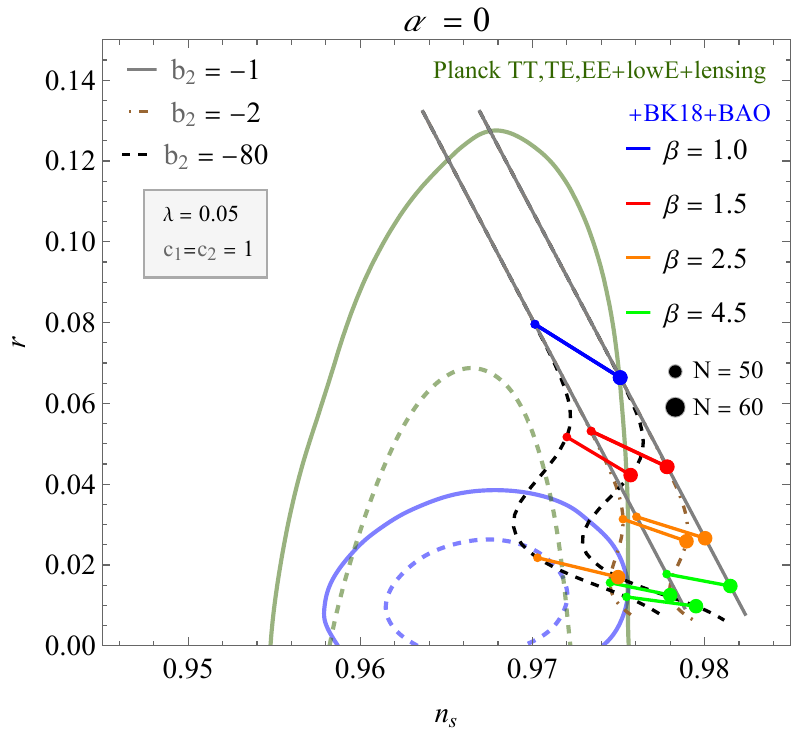}\label{plot_r_ns_1}}
\quad
\subfigure[]{\includegraphics[scale=0.6]{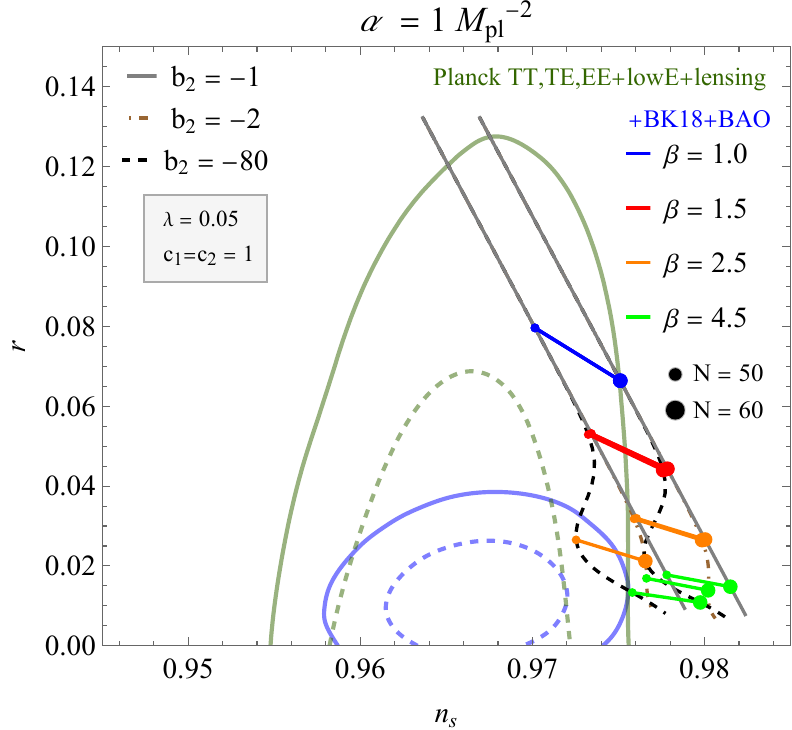}\label{plot_r_ns_2}}
\caption{\footnotesize{The $n_{s}-r$ plane for different values of $\beta$ and fixed $c_{1} = 1$ and $\lambda = 0.05$. The contours correspond to the BK18+BAO data.}}
\label{plot_r_ns_c_1}
\end{figure}

\begin{figure}[H]
	\centering
		\subfigure[]{\includegraphics[scale=0.6]{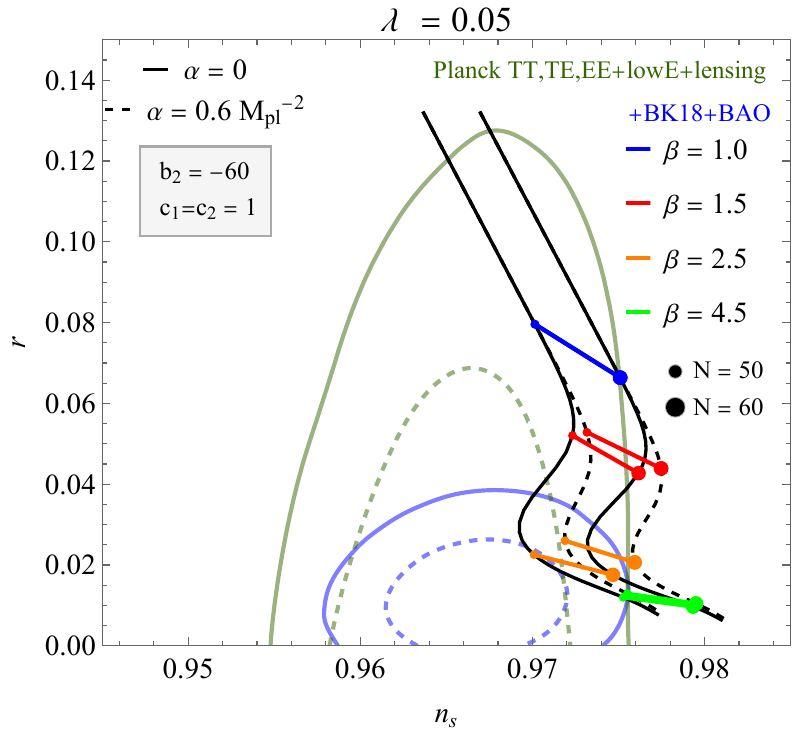}\label{plotlamb005bm60}}
			\quad
		\subfigure[]{\includegraphics[scale=0.6]{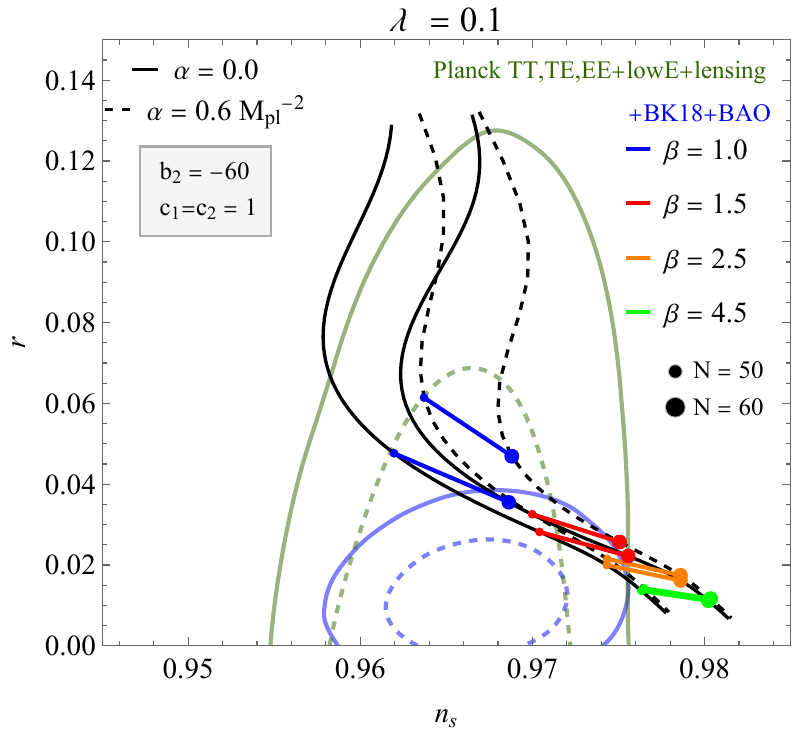}\label{plotlamb01bm60}}
		\caption{\footnotesize{The $n_{s}-r$ plane for different values of $\beta$ and $\alpha$ and keeping the values fixed $c_{2}=c_{1} = 1$ and $b_{2} = -60$. The contours correspond to the BK18+BAO data.}}
\label{plotc1bm60}
\end{figure}

\begin{figure}[H]
	\centering
		\subfigure[]{\includegraphics[scale=0.6]{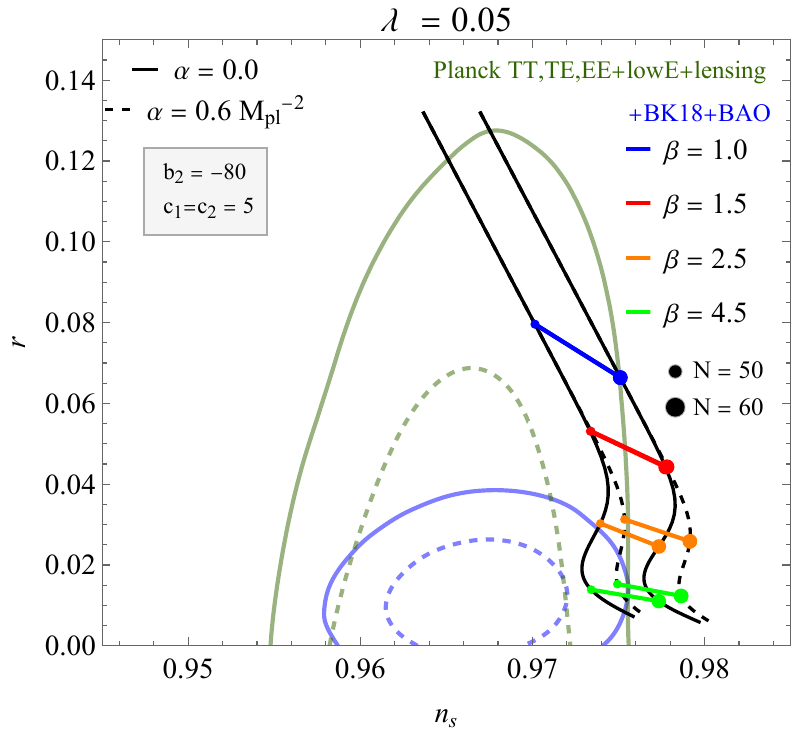}\label{plot_c5lamb005bm80}}
			\quad
		\subfigure[]{\includegraphics[scale=0.6]{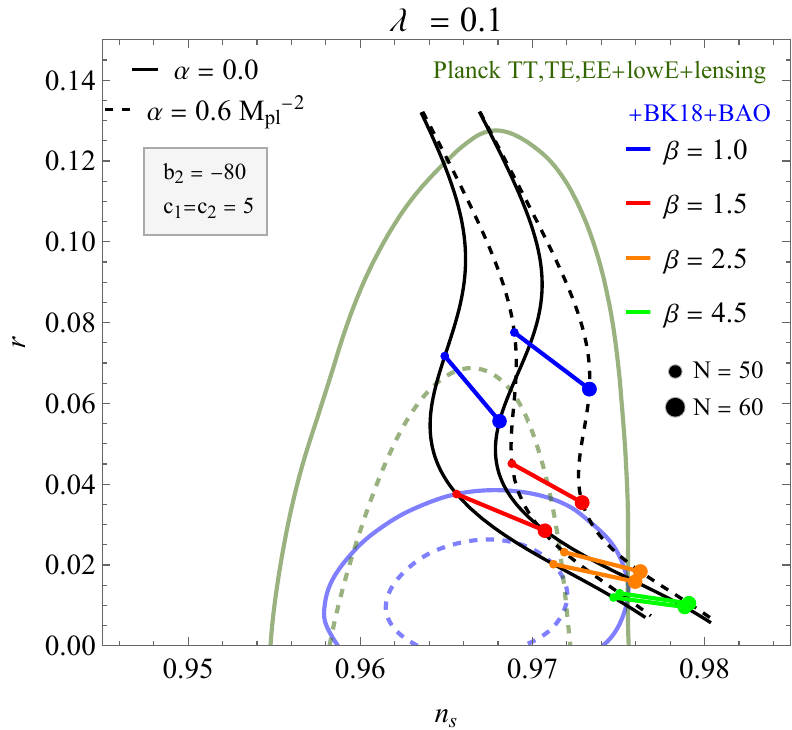}\label{plot_c5lamb01bm80}}
		\caption{\footnotesize{The $n_{s}-r$ plane for different values of $\beta$ and $\alpha$ keeping the values fixed $c_{1}=c_{2}=5$ and $b_{2} = -80$. The contours correspond to the BK18+BAO data.}}
\label{plotc5bm80}
\end{figure}

Finally, the spectral scalar index and the tensor-to-scalar ratio can be written in terms of $\phi_{i}$, as
{
\begin{equation}
n_s =1 - \dfrac{c_{1}^{2}\lambda^{2}}{2c_{2}^{2}(1+\alpha M_{pl}^{2})\left[1-\left(\dfrac{b_{2}}{c_{2}^{2}} + \dfrac{b_{1}}{c_{1}^{2}}\right)e^{-\phi_{i}/M_{pl}}\right]^{2}\left(1-\dfrac{\beta \lambda \phi_{e}}{M_{pl}}\right)^{2}}\left[1+2\beta - \left(\dfrac{1}{\lambda} + \beta(2 - \dfrac{\phi_{i}}{M_{pl}})\right)\left(\dfrac{b_{2}}{c_{2}^{2}} + \dfrac{b_{1}}{c_{1}^{2}}\right)e^{-\phi_{i}/M_{pl}}\right] 
\label{param_ns}
\end{equation}
\begin{equation}
r = \dfrac{4c_{1}^{2}\lambda^{2}}{c_{2}^{2}(1+\alpha M_{pl}^{2})\left[1-\left(\dfrac{b_{2}}{c_{2}^{2}} + \dfrac{b_{1}}{c_{1}^{2}}\right)e^{-\phi_{i}/M_{pl}}\right]\left(1-\dfrac{\beta \lambda \phi_{e}}{M_{pl}}\right)^{2}}
\label{param_r}
\end{equation}}
Similarly, as done for $\phi_{e}$ we can obtain the values of $\phi$ by solving \eqref{dil_N} numerically.
For $b_{2} = -b_{1}$, $c_{2}^{2} = c_{1}^{2}$ and $\alpha =0 $ the expressions $n_{s}$ and $r$ return to the standard $\beta$-exponential \cite{santos2018cmb}.
{In our work, we have $c_{1}=c_{2}$ while assuming $b_{1} = 1$. This way, in the figures \ref{plot_r_ns_1}, \ref{plotc1bm60} and \ref{plotc5bm80}, we check the results for the scalar index and the tensor-to-scalar ratio, in a plane $(n_{s}, r )$, considering two numbers for e-folds $N=50$ and $N=60$ and compared with the 68\% and 95\% contours obtained by Planck(2018)+BAO+BICEP/Keck Array data \cite{akrami2020planck, ade2021improved}.
In Fig. \ref{plot_r_ns_1} we see the influence of the parameter $b_{2}$ on the behavior of the curves, for $\lambda = 0.05$. For $b_2 =-1$ we have the result found in \cite{santos2018cmb} outside the regions delimited by +BK18+BAO data, however when we vary the parameter $b_2$ the curves lean towards this region. Comparing the figures \ref{plot_r_ns_1} and \ref{plot_r_ns_2}, we see that $\alpha$ contributes to attenuating the effect of $\beta$. In table \ref{tab1} we have the results for the scalar index and the tensor-to-scalar ratio for the values of $\beta$ bounded by the region of the +BK18+BAO data.}


{In figures \ref{plotc1bm60} and \ref{plotc5bm80} we have again the behavior of the curves for two values of e-foldings $N=50$ and $N=60$, we now test the contribution of the parameters $\alpha$ and $b_{2}$ with the change of $\lambda$ in the potential \eqref{potexp}.
In Fig. \ref{plotc1bm60} with $b_{2} = -60$ we have a reduction of the possible values of $\beta$ that includes the region delimited by the +BK18 + BAO data in $\lambda = 0.05$, compared with $\lambda = 0.1$. Furthermore, $\alpha$ reduces the approximation of the curves ($N=50$ and $N=60$) in the regions.
We saw in Fig. \ref{plot_r_ns_c_1} that $b_{2}$ contributes to the displacement of the curves to the left, however, note that in Fig \ref{plotc5bm80} where we have $c_{1}=c_{2}=5$ even for the parameter $b_{2} = -80$ the approximation of the curves is reduced, in both cases of $\lambda$, compared to the previous figure where $b_{2} = -60$. In this way, the influence of $c_{1}$ is similar to the $\alpha$ parameter, reducing the approximation of the curves to the delimited region of the data.
The results for the scalar index and the tensor-to-scalar ratio corresponding to the two figures mentioned above can be found in Table \ref{tab2} for values of $\beta$ bounded by the +BK18+BAO data. }{The fact that BAO tightens $r$ has been already noticed in several scenarios. See, for example, Ref.~ \cite{BICEP} for a large discussion on this issue.}


\begin{table}[ht!]
		\begin{center}
	\caption{\footnotesize{{  This table summarizes the $n_{s}-r$. We fix the e-folding number to be $N=50$, $c_{1}=c_{2}=1$ and $\lambda=0.05 $}}} 
		\label{tab1}	
			\begin{tabular}{|c||c|c|c|c|c|c|c|c|}
	\hline
	\multicolumn{1}{|c||}{\multirow{2}{*}{$ \beta $}}
	 & \multicolumn{2}{c|}{{\footnotesize $b_{2} = -1 \quad\text{and}\quad \alpha M_{pl}^{2} = 0$}}& \multicolumn{2}{c|}{{\footnotesize $b_{2} = -80 \quad\text{and}\quad \alpha M_{pl}^{2} = 0$}} &\multicolumn{2}{c|}{{\footnotesize $b_{2} = -1$ and  $\alpha M_{pl}^{2} = 1 $}} & \multicolumn{2}{c|}{{\footnotesize $b_{2} = -80 $ and  $\alpha M_{pl}^{2} = 1 $}}\\
	\cline{2-9} 
     & $n_{s}$ & $r$ & $n_{s}$ & $r$ & $n_{s}$ & $r$ & $n_{s}$ & $r$  \\
	\hline
  $2.2$  & 0.97551 & 0.036281 & 0.96907 & 0.027086  & 0.97551 & 0.036281 &  0.97277 & 0.032546  \\
  $2.7$  & 0.97634 & 0.029575 & 0.97123 & 0.019479  & 0.97634 & 0.029575 &  0.97275 & 0.023523  \\
  $3.0$  & 0.97671 & 0.026622 & 0.97247 & 0.017087  & 0.97671 & 0.026622 &  0.97335 & 0.020223  \\
  $3.6$  & 0.97725 & 0.022191 & 0.97414 & 0.014311  & 0.97725 & 0.022191 &  0.97459 & 0.016314  \\
  $4.1$  & 0.97759 & 0.019488 & 0.97499 & 0.012928  & 0.97759 & 0.019488 &  0.97535 & 0.014426   \\
	\hline
 			\end{tabular}
		\end{center}
\end{table} 

\begin{table}[ht!]
		\begin{center}
	\caption{\footnotesize{{  This table summarizes the $n_{s}-r$. We fix the e-folding number to be $N=50$}}} 
		\label{tab2}	
			\begin{tabular}{c|c|c|c|c|c|c|c|c|}
	\cline{2-9}
	&\multicolumn{4}{c|}{{\footnotesize $\lambda = 0.05,\quad c_{1}=c_{2} = 1 \quad \text{and}\quad  b_{2} = -60 $}}&\multicolumn{4}{c|}{{\footnotesize $\lambda = 0.05,\quad c_{1}=c_{2} = 1 \quad \text{and}\quad  b_{2} = -80 $}}\\
	\hline
	\multicolumn{1}{|c||}{\multirow{2}{*}{$ \beta $}}
	 & \multicolumn{2}{c|}{{\footnotesize $\alpha M_{pl}^{2} = 0 $}}& \multicolumn{2}{c|}{{\footnotesize $\alpha M_{pl}^{2} = 0.6 $}} &\multicolumn{2}{c|}{{\footnotesize $\alpha M_{pl}^{2} = 0 $}} & \multicolumn{2}{c|}{{\footnotesize $\alpha M_{pl}^{2} = 0.6 $}}\\
	\cline{2-9} 
    \multicolumn{1}{|c||}{ }  & $n_{s}$ &   $r$ &  $n_{s}$ &   $r$ & $n_{s}$ &   $r$ &  $n_{s}$ &   $r$  \\
	\hline
   \multicolumn{1}{|c||}{$2.0$}  & 0.96950 & 0.033537 &  0.97274 & 0.037130  & 0.97440 & 0.039341 &  0.97489 & 0.039717 \\
   \multicolumn{1}{|c||}{$2.5$}  & 0.97009 & 0.022558 &  0.97189 & 0.026011  & 0.97395 & 0.030277 &  0.97535 & 0.031291 \\
   \multicolumn{1}{|c||}{$3.0$}  & 0.97212 & 0.017431 &  0.97282 & 0.019737  & 0.97317 & 0.023815 &  0.97522 & 0.025328 \\
   \multicolumn{1}{|c||}{$4.5$}  & 0.97527 & 0.012000 &  0.97551 & 0.012902  & 0.97341 & 0.013908 &  0.97494 & 0.015281 \\
	\hline
	\cline{2-9}
	&\multicolumn{4}{c|}{{\footnotesize $\lambda = 0.1, \quad c_{1}=c_{2} = 5 \quad \text{and}\quad  b_{2} = -60 $}}&\multicolumn{4}{c|}{{\footnotesize $\lambda = 0.1,\quad c_{1}=c_{2} = 5 \quad \text{and}\quad  b_{2} = -80 $}}\\
	\hline
	\multicolumn{1}{|c||}{\multirow{2}{*}{$ \beta $}}
	 & \multicolumn{2}{c|}{{\footnotesize $\alpha M_{pl}^{2} = 0 $}}& \multicolumn{2}{c|}{{\footnotesize $\alpha M_{pl}^{2} = 0.6 $}} &\multicolumn{2}{c|}{{\footnotesize $\alpha M_{pl}^{2} = 0 $}} & \multicolumn{2}{c|}{{\footnotesize $\alpha M_{pl}^{2} = 0.6 $}}\\
	\cline{2-9} 
    \multicolumn{1}{|c||}{ }  & $n_{s}$ &   $r$ &  $n_{s}$ &   $r$ & $n_{s}$ &   $r$ &  $n_{s}$ &   $r$  \\
	\hline
   \multicolumn{1}{|c||}{$1.2$}  & 0.96657 & 0.035820 &  0.96629 & 0.044100 & 0.96406 & 0.053499 &  0.96896 & 0.061721  \\
   \multicolumn{1}{|c||}{$2.6$}  & 0.97449 & 0.019523 &  0.97454 & 0.020953 & 0.97155 & 0.019428 &  0.97213 & 0.022190  \\
   \multicolumn{1}{|c||}{$3.2$}  & 0.97529 & 0.017192 &  0.97539 & 0.018228 & 0.97303 & 0.016015 &  0.97345 & 0.017890  \\
   \multicolumn{1}{|c||}{$4.0$}  & 0.97600 & 0.014816 &  0.97615 & 0.015552 & 0.97421 & 0.013200 &  0.97458 & 0.014474  \\
	\hline		
 			\end{tabular}
		\end{center}
\end{table} 

\begin{figure}[H]
	\centering
		\subfigure{\includegraphics[scale=0.6]{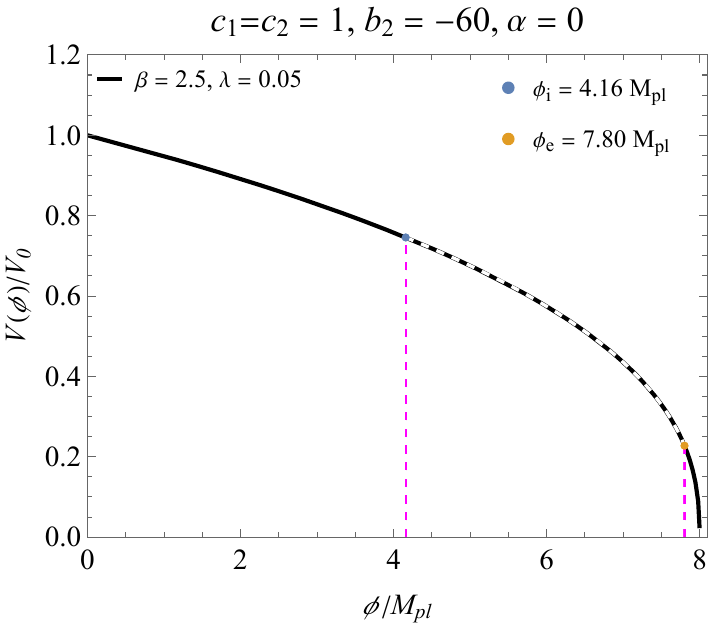}}
			\quad
		\subfigure{\includegraphics[scale=0.6]{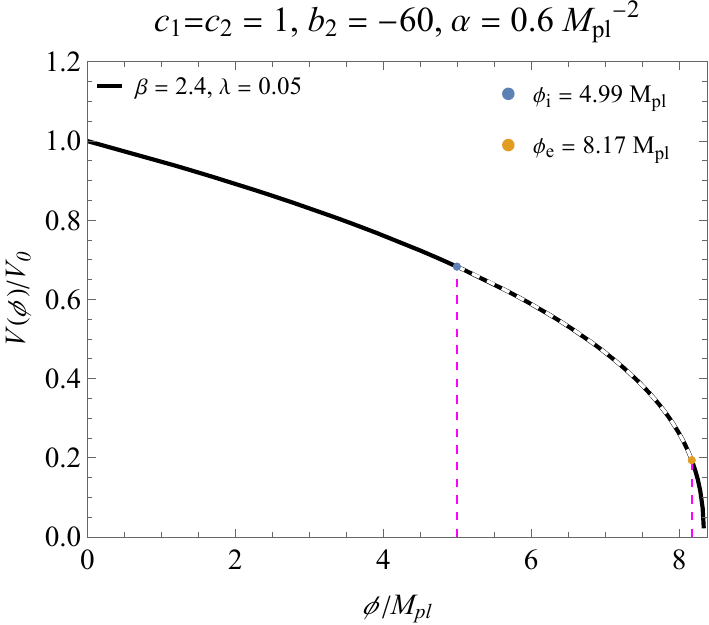}}
			\quad
		\subfigure{\includegraphics[scale=0.6]{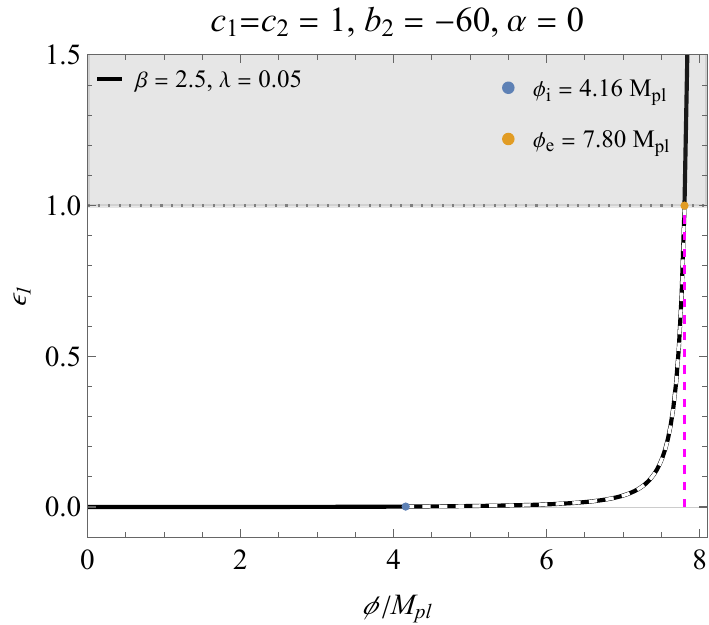}}
			\quad
		\subfigure{\includegraphics[scale=0.6]{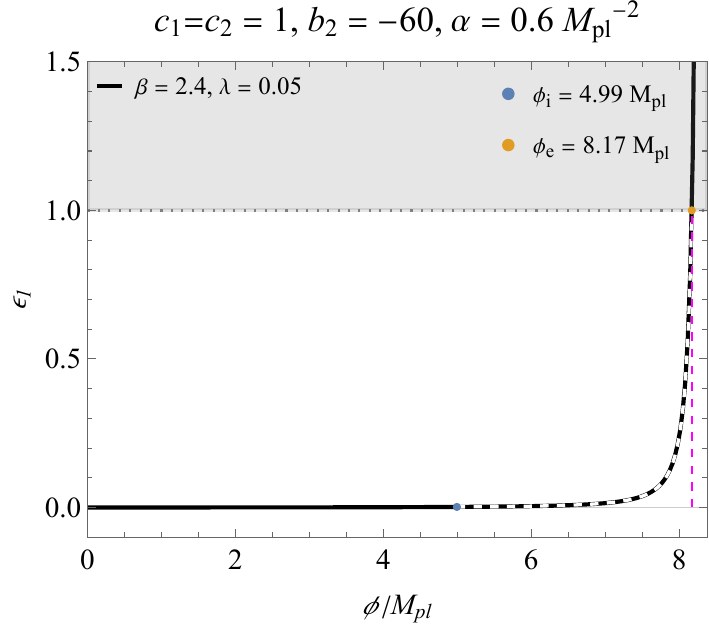}}
		\caption{\footnotesize{Top graphs represent the behavior of the potential for the parameters $b_{2} = -60$ and $\lambda = 0.05$, within the inflation regime, corresponding to the results presented in Fig. \ref{plotlamb005bm60}. Bottom panels are the respective slow-roll parameters $\epsilon_{1}$ as a function of $\phi$.}}
		\label{V_epson_bm60_lamb005}
\end{figure}

\begin{figure}[H]
	\centering
		\subfigure{\includegraphics[scale=0.6]{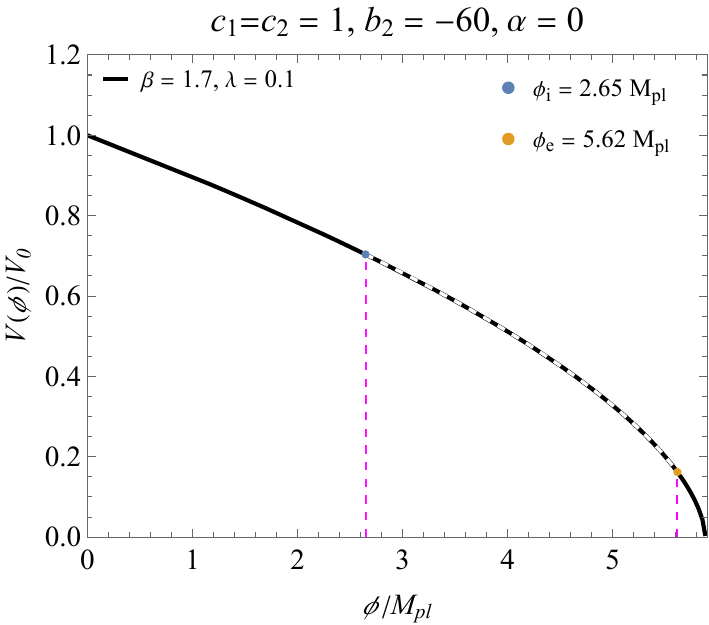}}
			\quad
		\subfigure{\includegraphics[scale=0.6]{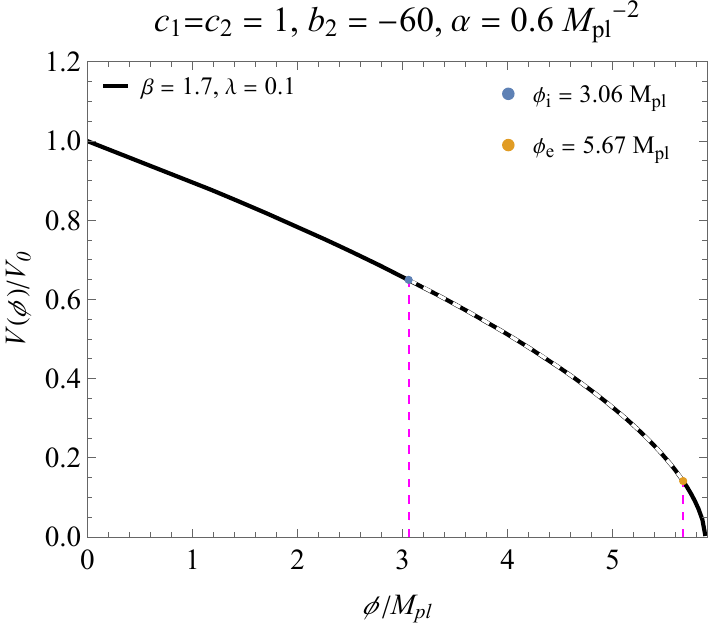}}
			\quad
		\subfigure{\includegraphics[scale=0.6]{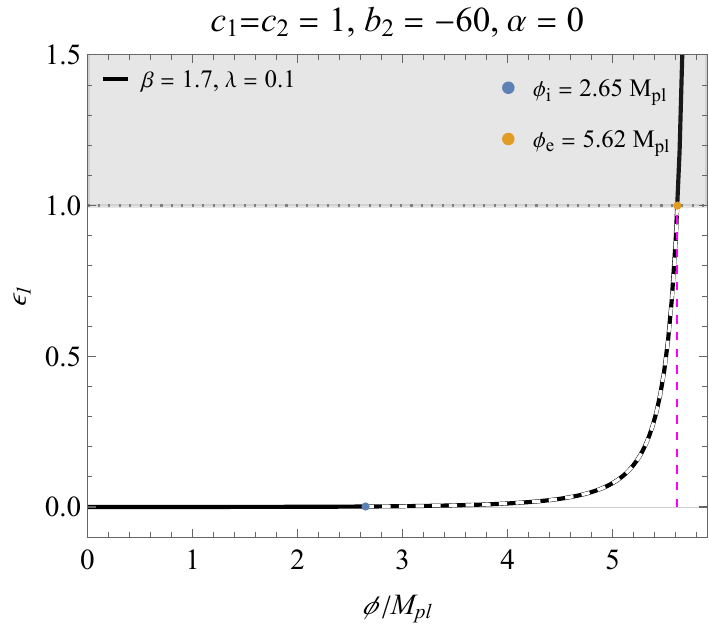}}
			\quad
		\subfigure{\includegraphics[scale=0.6]{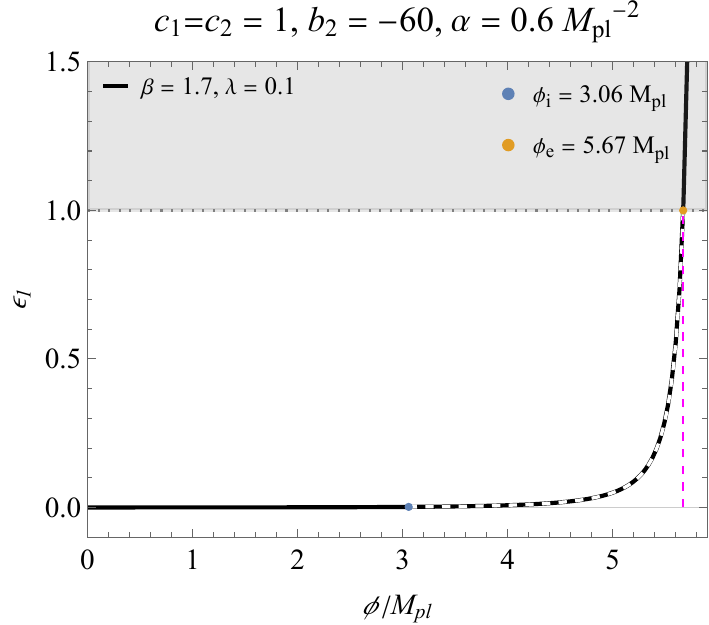}}
		\caption{\footnotesize{Top graphs represent the behavior of the potential for the parameters $b_{2} = -60$ and $\lambda = 0.1$, within the inflation regime, corresponding to the results presented in Fig. \ref{plotlamb01bm60}. Bottom panels are the respective slow-roll parameters $\epsilon_{1}$ as a function of $\phi$.}}
		\label{V_epson_bm60_lamb01}
\end{figure}

\begin{figure}[H]
	\centering
		\subfigure{\includegraphics[scale=0.6]{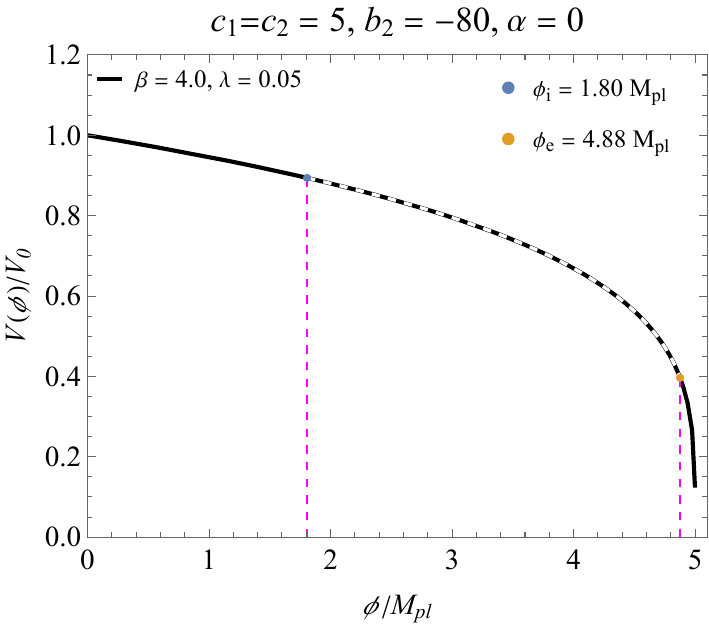}}
			\quad
		\subfigure{\includegraphics[scale=0.6]{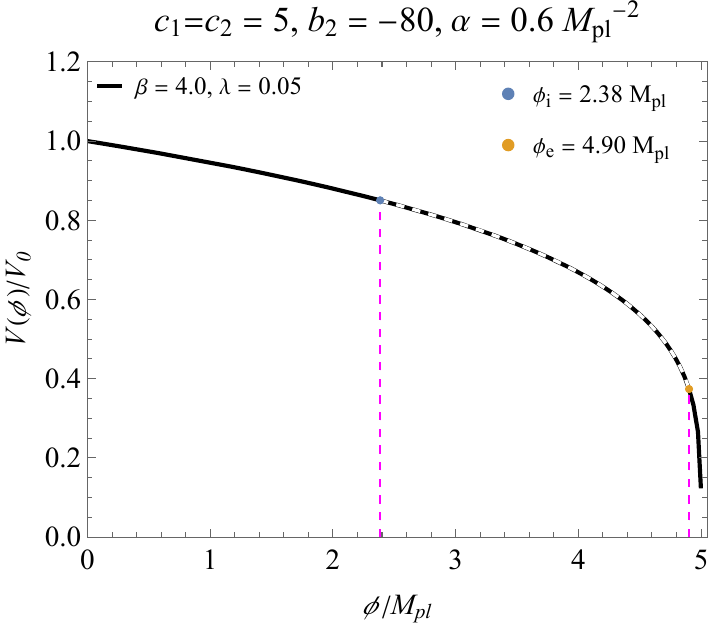}}
			\quad		
		\subfigure{\includegraphics[scale=0.6]{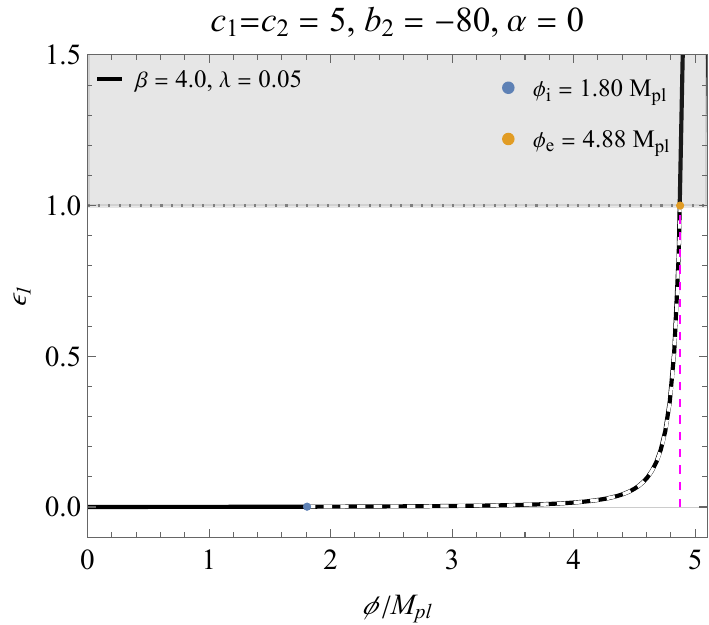}}
			\quad
		\subfigure{\includegraphics[scale=0.6]{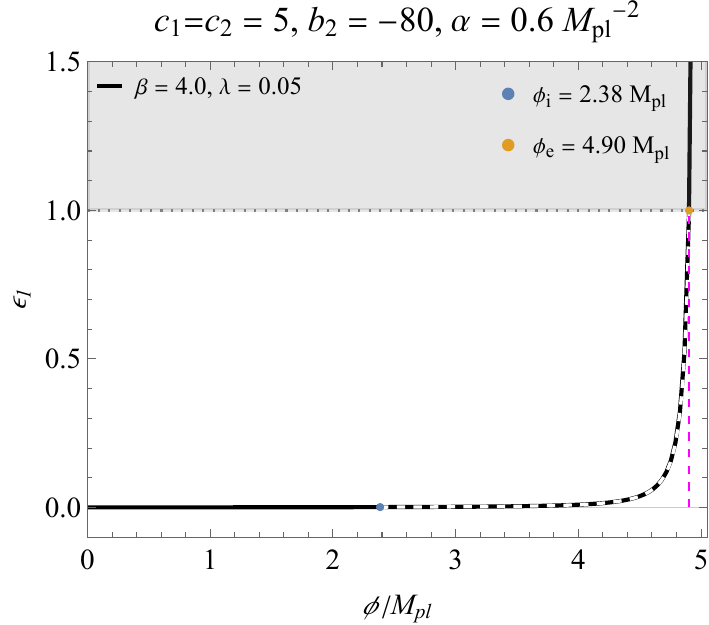}}
		\caption{\footnotesize{Top graphs represent the behavior of the potential for the parameters $b_{2} = -80$ and $\lambda = 0.05$, within the inflation regime, corresponding to the results presented in Fig. \ref{plot_c5lamb005bm80}. Bottom panels are the respective slow-roll parameters $\epsilon_{1}$ as a function of $\phi$.}}
		\label{V_epson_bm80_lamb005}
\end{figure}

\begin{figure}[H]
	\centering
		\subfigure{\includegraphics[scale=0.6]{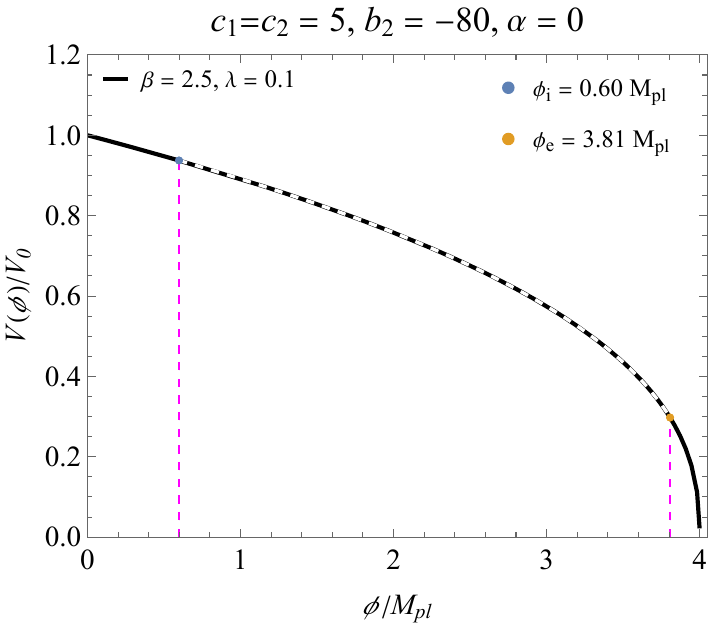}}
			\quad
		\subfigure{\includegraphics[scale=0.6]{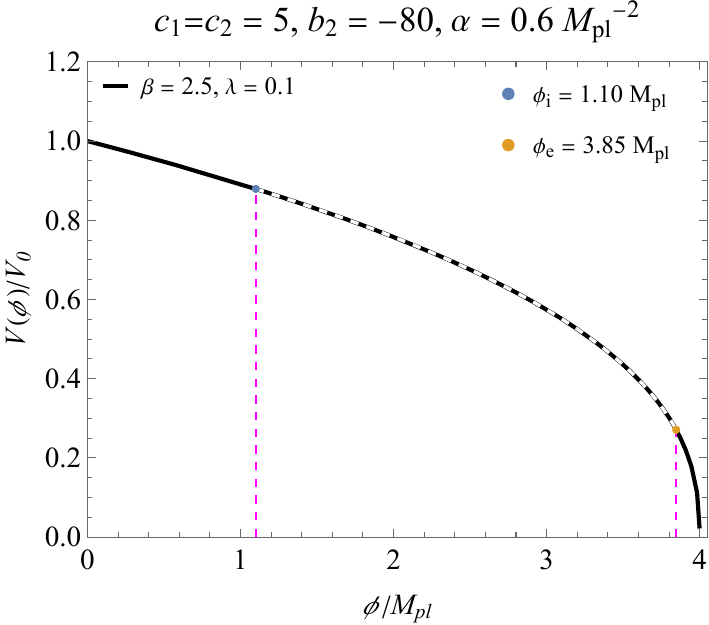}}
			\quad
		\subfigure{\includegraphics[scale=0.6]{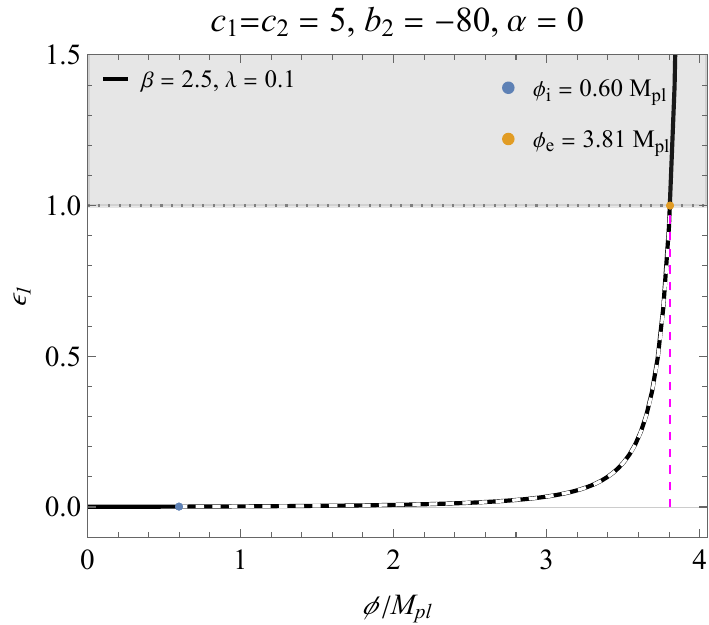}}
			\quad
		\subfigure{\includegraphics[scale=0.6]{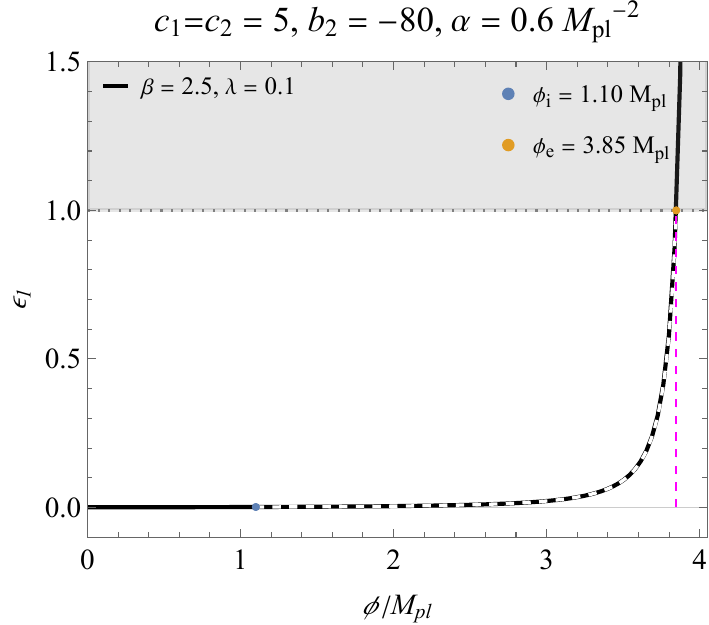}}
		\caption{\footnotesize{Top graphs represent the behavior of the potential for the parameters $b_{2} = -80$ and $\lambda = 0.1$, within the inflation regime, corresponding to the results presented in Fig. \ref{plot_c5lamb01bm80}. Bottom panels are the respective slow-roll parameters $\epsilon_{1}$ as a function of $\phi$.}}
		\label{V_epson_bm80_lamb01}
\end{figure}

{In the following figures, we have the behavior of the potential \eqref{potexp} for the parameters shown in figures \ref{plotc1bm60} and \ref{plotc5bm80}, checking the beginning and end of inflation. We choose an appropriate value for $\beta$ within the region delimited by the BK18+BAO data. In the bottom panels, we have the slow-roll parameter $\epsilon_{1}$ as a function of $\phi$ corresponding to each potential, and the shaded area indicates the breakdown of the slow-roll inflation. The figures \ref{V_epson_bm60_lamb005} and \ref{V_epson_bm60_lamb01} correspond to the cases shown in Fig \ref{plotc1bm60}, while the figures \ref{V_epson_bm80_lamb005} and \ref{V_epson_bm80_lamb01} correspond to Fig \ref{plotc5bm80}. It is possible to notice that the contribution of $\alpha$ reduces the inflation band.}

 
\begin{figure}[H]
	\centering
		\subfigure[]{\includegraphics[scale=0.6]{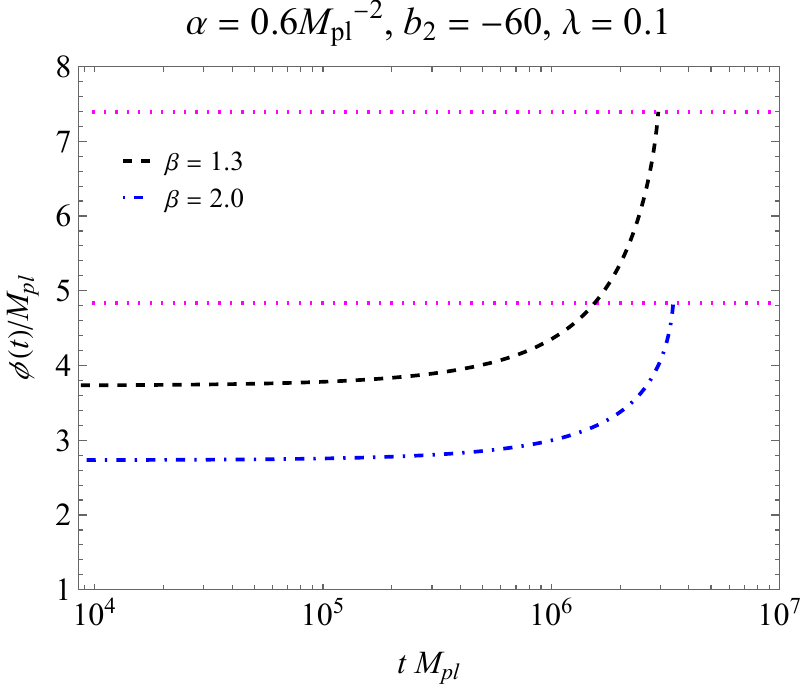}\label{phi_t_c1bm60}}
			\quad
		\subfigure[]{\includegraphics[scale=0.6]{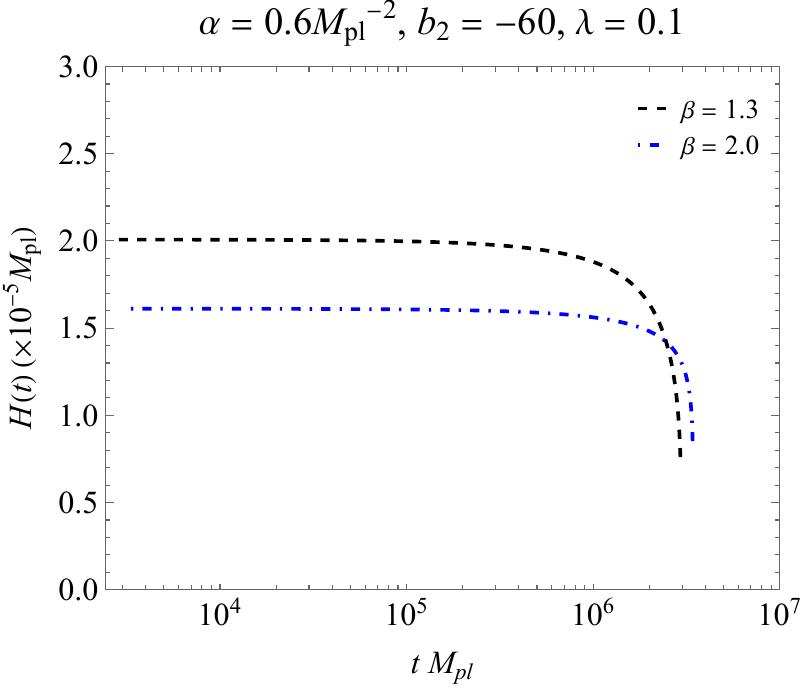}\label{H_t_c1bm60}}
			\quad
		\subfigure[]{\includegraphics[scale=0.6]{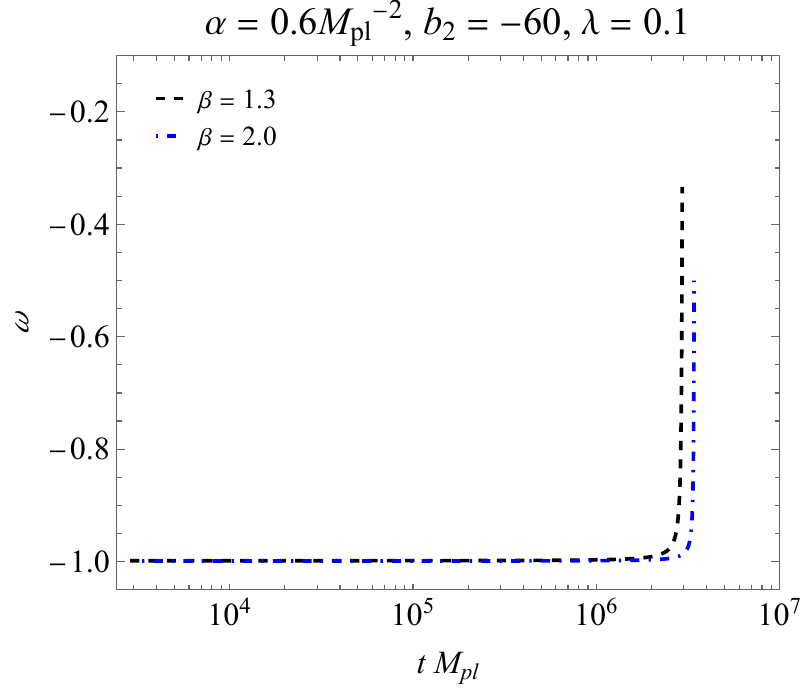}\label{W_t_c1bm60}}
		\caption{\footnotesize{The figure shows the behavior of $\phi(t)$, $H(t)$ and $\omega(t)$ for the fixed values of $b_{2}=-60$, $\alpha=0.6$ and $c_{1}=c_{2} = 1$. In the graph (a) we have the field $\phi$ as a function of time. The horizontal dotted lines delimit the end of inflation. In (b) we have $H(t)$ in the order of Planck mass. For the end of inflation we have $H_{e} = 7.43\times 10^{-6}M_{pl}$ for $\beta = 1.3$ and $H_{e} = 8.34\times 10^{-6}M_{pl}$ for $\beta = 2.0$. In (c) we have the behavior of the equation of state.}}
		\label{phi_H_bm60_lamb01}
\end{figure}

\begin{figure}[H]
	\centering
		\subfigure{\includegraphics[scale=0.6]{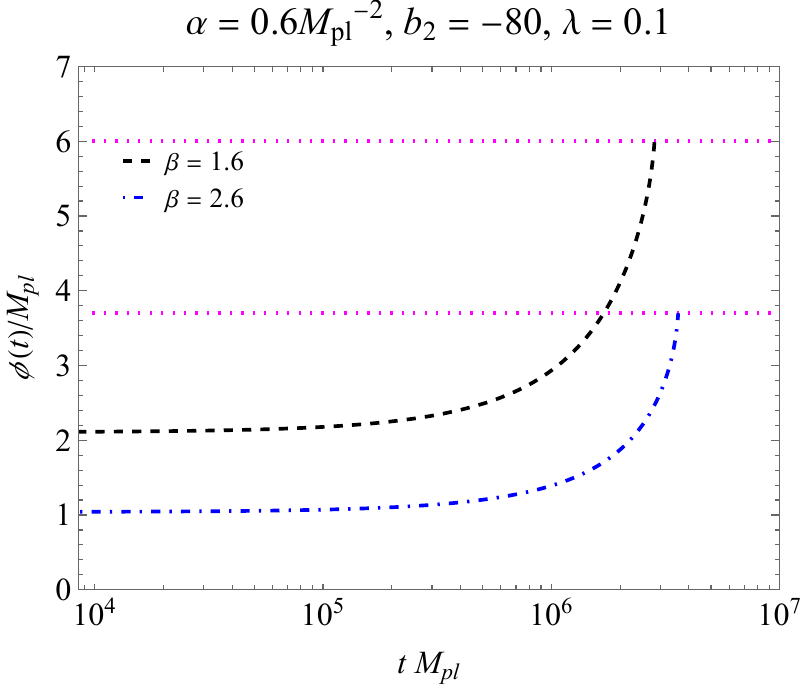}\label{phi_t_c5bm80}}
			\quad
		\subfigure{\includegraphics[scale=0.6]{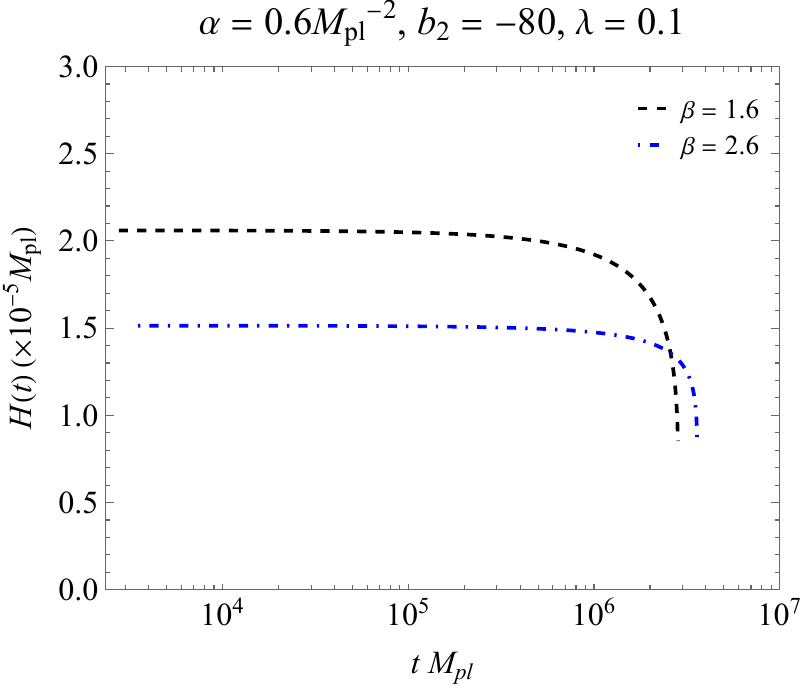}\label{H_t_c5bm80}}
			\quad
		\subfigure{\includegraphics[scale=0.6]{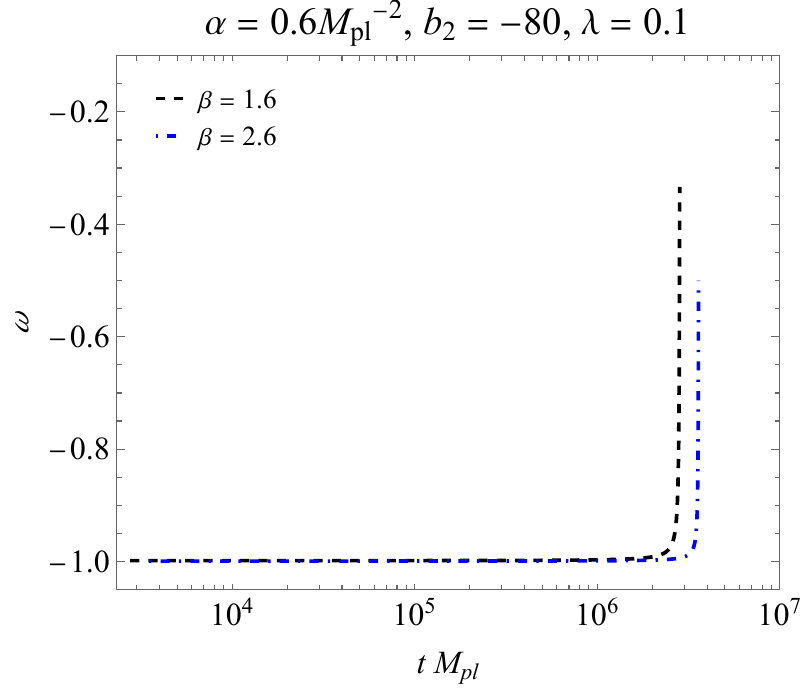}\label{W_t_c5bm80}}
		\caption{\footnotesize{The figure shows the behavior of $\phi(t)$, $H(t)$ and $\omega(t)$ for the fixed values of $b_{2}=-80$, $\alpha=0.6$ and $c_{1}=c_{2} = 5$. In the graph (a) we have the field $\phi$ as a function of time. In (b) we have $H(t)$ at the end of inflation, i.e.,  $H_{e} = 8.53\times 10^{-6}M_{pl}$ for $\beta = 1.6$ and $H_{e} = 8.58\times 10^{-6}M_{pl}$ for $\beta = 2.6$. In (c) we have the behavior of the equation of state.}}
		\label{phi_H_bm80_lamb01}
\end{figure}

{For instantaneous reheating in which one finds  maximal temperature, we can further estimate the reheating temperature by assuming the energy density $\rho_{e} = (1+\alpha M_{pl}^{2})V(\phi_{e})$, so from the Stefan–Boltzmann law the reheating temperature is given by
\begin{eqnarray}
T_{rh} = \left(\dfrac{30 \rho}{\pi^{2}g_{*}}\right)^{1/4} \cong \left(\dfrac{30(1+\alpha M_{pl}^{2})V_{0}\left(1-\beta\lambda\phi_{e}/M_{pl}\right)^{1/\beta}}{\pi^{2}g_{*}}\right)^{1/4},
\label{Tr}
\end{eqnarray} 
where $g_{*} \sim 106$ are the degrees of freedom of the Standard Model.
The results for the reheating temperature can be found in Table \ref{tab3}.
{In figures \ref{phi_H_bm60_lamb01} and \ref{phi_H_bm80_lamb01} we have the behavior of the parameters $H(t)$, $\phi(t)$ and $\omega$.} {Notice the approximately de Sitter regime as $H\approx const.$ before the end of inflation}}. For the field $\phi(t)$ in figures \ref{phi_t_c1bm60} and \ref{phi_t_c5bm80} the behavior of the curves is limited by the horizontal dotted line, which shows the end of inflation in each scenario. 
In Figures \ref{H_t_c1bm60} and \ref{H_t_c5bm80} we have the behavior of $H(t)$ on the scale $\times 10^{-5}M_{pl}$, we verify that in Fig. \ref{phi_H_bm60_lamb01} for the end of inflation $H_{e} = 7.43\times 10^{-6}M_{pl}$ with $\beta = 1.3$ and $H_{e} = 8.34\times 10^{-6}M_{pl}$ with $\beta = 2.0$ using the equation \eqref{Tr} we have the respective values of the reheating temperature $T_{rh} = 3.59 \times 10^{15}GeV$ and $T_{rh} = 3.80 \times 10^{15}GeV$.
On the other hand, for Fig. \ref{phi_H_bm60_lamb01} we have $H_{e} = 7.43\times 10^{-6}M_{pl}$ with $\beta = 1.3$ and $H_{e} = 8.34\times 10^{-6}M_{pl}$ with $\beta = 2.0$, for the respective temperature values $T_{rh} = 3.85 \times 10^{15}GeV$ and $T_{rh} =3.86 \times 10^{15}GeV$. Finally, we have in figures \ref{W_t_c1bm60} and \ref{W_t_c5bm80} the behavior for the effective equation of state, $\omega = -1-2\dot{H}/3H^{2}$.


\begin{table}[ht!]
		\begin{center}
	\caption{\footnotesize{Values for reheating temperature $T_{rh}$ in ($GeV$) for $\lambda=0.1$ and $c_{1}=c_{2}=1$}} 
		\label{tab3}	
			\begin{tabular}{c||c|c|c|c}
	\hline
	\multicolumn{1}{c||}{\multirow{2}{*}{$ \beta $}}
	 & \multicolumn{2}{c|}{ $b_{2} = -60 $}& \multicolumn{2}{c}{ $b_{2} = -80 $} \\
	\cline{2-5} 
     & \qquad $\alpha M_{pl}^{2} = 0$ \qquad & \qquad $\alpha M_{pl}^{2} = 1$ \qquad & \qquad$\alpha M_{pl}^{2} = 0$ \qquad &\qquad $\alpha M_{pl}^{2} = 1$ \qquad  \\
	\hline
  $1.4$  & $3.55324 \times 10^{15}$ & $3.67690 \times 10^{15}$ & $3.56277 \times 10^{15}$ & $3.67147 \times 10^{15}$   \\
  $2.8$  & $3.84265 \times 10^{15}$ & $3.90058 \times 10^{15}$ & $3.85847 \times 10^{15}$ & $3.91257 \times 10^{15}$   \\
  $4.2$  & $3.82876 \times 10^{15}$ & $3.87660 \times 10^{15}$ & $3.84495 \times 10^{15}$ & $3.88967 \times 10^{15}$  \\
\hline
 			\end{tabular}
		\end{center}
\end{table}

\section{Conclusions}\label{VI}

Our results show how to rescue models in the context of dilatonic $f(R,T)$ gravity in view of the cosmological observations of the Planck 2018 Collaboration. {By considering a specific range of parameters adopted in the present analysis, as can be seen in the previous section along with  the figures and  tables, we check the results for the scalar index and the tensor-to-scalar ratio, in a plane $(n_{s}, r )$, by considering two numbers for e-folds $N=50$ and $N=60$ and compare with the 68\% and 95\% contours obtained by Planck(2018)+BAO+BICEP/Keck Array data \cite{akrami2020planck, ade2021improved}. The best results, i.e., those ones that fall inside these contours are related to $\lambda=0.05$, $\beta=2.5$ and $b_2=-60$ ($\alpha=0$) or $b_2=-80$ ($\alpha=M_{pl}^{-2}$). For these values of $b_2$, $H\approx const.$, that means the Universe evolves in the approximately de Sitter regime before the end of inflation.} As we well know many interesting models in the past were ruled out by updated data along the time according to revisions of cosmological observational data. In the present study we turn our attention to the well-explored $\beta$-exponential. This model, {that in our setup corresponds to $b_2=-1$}, suffers strong restrictions from the Planck 2018 data. However, it was soon rescued in the context of gravity non-minimally coupled to scalar field \cite{dosSantos:2021vis}. In the present investigation we reviewed this problem in the context of dilatonic $f(R,T)$ gravity. We show that the presence of the dilaton coupling in the intermediate regime plays a fundamental role in rescuing the model. {Finally, the model presents a fairly good approximate reheating temperature. In our present scope, we just estimate the instantaneous reheating which leads to maximal temperature. A study in the future concerns to extend the present analysis to consider the decay of the inflaton in several species due to the coupling with other fields.}

\acknowledgments

We would like to thank CNPq, CAPES and FAPESQ-PB. J.A.V. Campos would like to thank FAPESQ-PB/CNPq
n$^0$ 77/2022 for financial support and F.A. Brito acknowledges CNPq  (Grant no.
309092/2022-1). C.H.A.B. Borges also acknowledges PIQIFPB for financial support.

\end{document}